\documentclass[aps,pre,superscriptaddress,twocolumn,showpacs]{revtex4}

\usepackage[dvips]{graphicx}
\usepackage{amsmath}
\usepackage{amssymb}
\usepackage{natbib}
\usepackage{psfrag}
\usepackage{color}
\usepackage{epsfig}
\usepackage{subfigure}

\newcommand{\rev}[1]{\textcolor{black}{#1}}
\newcommand{\secondrev}[1]{\textcolor{black}{#1}}

\begin{document}

\title{\rev{Growth-rate-dependent dynamics of a bacterial genetic oscillator}}

\author{Matteo Osella}
\email[correspondence to: ]{matteo.osella@upmc.fr}

\affiliation{Genomic Physics Group, UMR7238 CNRS ``Microorganism
  Genomics''}

\affiliation{University Pierre et Marie Curie, 15, rue de l'\'{E}cole
  de M\'{e}decine, Paris, France}

\author{Marco Cosentino Lagomarsino} 

\affiliation{Genomic Physics Group, UMR7238 CNRS ``Microorganism
  Genomics''}

\affiliation{University Pierre et Marie Curie, 15, rue de l'\'{E}cole
  de M\'{e}decine, Paris, France}

\affiliation{Dipartimento di Fisica, University of Torino, Via
  P. Giuria 1, Torino, Italy}

\date{\today}

\begin{abstract}
  Gene networks exhibiting oscillatory dynamics are widespread in
  biology. The minimal regulatory designs giving rise to oscillations
  have been implemented synthetically and studied by mathematical
  modeling.  However, most of the available analyses generally neglect
  the coupling of regulatory circuits with the cellular ``chassis'' in
  which the circuits are embedded. For example, the 
 intracellular macromolecular composition of fast-growing
  bacteria changes with growth rate. As a consequence, important
  parameters of gene expression, such as ribosome concentration or
  cell volume, are growth-rate dependent, ultimately coupling the
  dynamics of genetic circuits with cell physiology. This work
  addresses the effects of growth rate on the dynamics of a
  paradigmatic example of genetic oscillator, the
  repressilator. Making use of empirical growth-rate dependences of
  parameters in bacteria, we show that the repressilator dynamics can
  switch between oscillations and convergence to a fixed point
  depending on the cellular state of growth, and thus on the 
  nutrients it is fed.  The physical support of the
  circuit (type of plasmid or gene positions on the chromosome) also
  plays an important role in determining the oscillation stability and
  the growth-rate dependence of period and amplitude.  This analysis
  has potential application in the field of synthetic biology, and
  suggests that the coupling between endogenous genetic oscillators
  and cell physiology can have substantial consequences for 
  their functionality.
\end{abstract}

\maketitle

\section{Introduction}

Oscillatory behavior is widespread and fundamentally important in
biological systems, from circadian clocks to cell-cycle
control~\cite{Goldbeter1997}.  At the level of a single cell,
oscillations can be sustained by regulatory circuits, the basic
components of which are nucleic acids, proteins, and biochemical
interactions~\cite{Goldbeter2002,Tyson2008}.
However, naturally occurring regulatory systems are complex. Thus, our
path to rationalize their behavior has to pass through simplifications
and basic underlying principles~\cite{Purcell2010}.  The study of
minimal circuits, by mathematical modeling and experimental synthetic
engineering, allows a quantitative approach to \rev{the study of} gene
expression~\cite{Bintu2005,Alon2006,Wall2004,Li2011}.

This simplifying approach is not only useful for studying the extant
regulatory circuits. Understanding the basic principles and dynamical
properties of regulatory networks makes it possible to design and
produce synthetic circuits that can perform specific functions in a
predictable
manner~\cite{Andrianantoandro2006,Guido2006,Haseloff2009,Nandagopal2011},
with potentially relevant future biotechnological and biomedical
applications~\cite{Haseloff2009,Weber2012}.  A paradigmatic example of
a synthetic genetic oscillator, and one of the earliest \textit{in
  vivo} realizations~\cite{Elowitz2000}, is the so-called
``repressilator'', a three-gene cyclic circuit where each gene
protein-product represses the synthesis of its
successor~(Fig.~\ref{fig1}A).

From a physics standpoint \rev{it is desirable to build minimal, but
  effective and testable models, with qualitative and possibly
  quantitative predictive power for experiments. To this end,} dealing
with \rev{controlled}, well-characterized, and isolated systems is
necessary.
Unfortunately, every genetic circuit, endogenous or synthetically
implemented in a living cell, cannot be truly considered isolated from
cellular processes. These processes are strongly affected by the
physiological state of the cell~\cite{Klumpp2009}, and thus by the
environmental conditions and type/availability of nutrients.  For
example, the macromolecular composition of bacterial cells in ``steady
exponential'' growth \rev{(i.e. constantly dividing at the same
  rate~\footnote{\rev{This state is conventionally referred to as
      ``steady'' because its maintenance requires that a nutrient sink
      (the cells) is constantly replenished with a source. }})}
changes substantially with growth rate (the rate of cell
proliferation)~\cite{Bremer1996,Scott2010,Scott2011}. Importantly, in
this case, growth rate appears to \rev{encapsulate} most of the
physiological changes. \rev{In other words, cells growing on different
  nutrients but at similar growth rates are in many ways equivalent,
  considering important global cell parameters such as concentrations
  of ribosomes and RNA polymerases (the molecular machines effecting
  translation and transcription respectively).}  \rev{ Such
  growth-rate dependent parameters affect gene expression, coupling
  its dynamics } with the cell state, with relevant consequences for
genetic circuit functioning~\cite{Klumpp2009,Tan2009,Kwok2010}.

\rev{The emergent quantitative ``laws'' of bacterial physiology,
  linking cell composition and growth rate, are reminiscent of those
  of thermodynamics, and were the subject of recent advances using
  physical modeling~\cite{Scott2010,Scott2011}. Previously, they were
  in part captured by early studies in the 1950's and
  60's~\cite{Bremer1996}. However, to date, they are not completely
  characterized.}
For \textit{E.~coli}, the best studied bacterial species, the growth
rate dependences of various cellular parameters were evaluated
\rev{phenomenologically} in a study by Klumpp and
coworkers~\cite{Klumpp2009}, compiling results from multiple
experiments. Leveraging on these empirical data, they analyzed the
growth-rate dependence of the steady state \rev{protein concentration} of a
constitutively expressed (i.e. unregulated) gene and few other simple
genetic circuits, such as the self-regulator and the toggle
switch. Their modeling strategy will be briefly reviewed in
section~\ref{growth-rate-dep}.
Beyond steady state, knowing the empirical growth-rate dependence of
gene expression parameters allows in principle to revisit the
dynamical properties of genetic circuits, introducing the
physiological cell state as a new player.

This work addresses  the growth-rate dependence of a
biological oscillator dynamics, and considers the effect of cell state
on a repressilator, integrated on either a plasmid or the
\textit{E.~coli} chromosome. \rev{From the physical modeling
  viewpoint, this entails understanding the models and
  phenomenological laws that relate the circuit parameters to the cell
  physiological state. Incorporating these features in otherwise
  fairly standard models leads to predictions that are
  experimentally relevant.}
After reviewing the conditions for oscillations and the repressilator
dynamics at fixed-growth rate (and extending some known results), we
will show how the physiology of the cell can alter qualitatively and
quantitatively the dynamical properties of the system. The main result
is that the conditions for stable oscillations, as well as their
amplitude and period are growth-rate dependent. This implies that a
genetic oscillator can display distinct dynamical behaviors in
different environments and nutrient conditions.  Additionally, the
circuit \rev{support}, e.g. type of plasmid or chromosomal position of
the genes, also contributes to its dynamics, through gene
dosage. Specifically, the range of growth rates in which oscillations
are observable will vary with \rev{support}, in a manner that could be
exploited both biologically and technologically.  Our predictions,
although based on a simplified model, are experimentally testable in a
straightforward way \rev{on synthetically realized repressillator
  circuits, and more in general set a framework for the more complex
  case of endogenous oscillators in bacteria, prominently the cell
  cycle and circadian clocks~\cite{Lenz2011}}.

The paper is organized as follows. Section~\ref{model} reviews the
modeling strategy adopted to include the growth rate as a variable in
gene expression (Subsection~\ref{growth-rate-dep}) and explains how
this strategy can be applied to a mathematical description of the
repressilator (Subsection~\ref{repressilator_intro}).
Appendix~\ref{app1} reviews the Cooper-Helmstetter
model~\cite{Cooper1968}, an empirical model of DNA replication in
fast-growing bacteria that is an essential part of our approach, and
justifies in more detail some of the model assumptions presented in
Section~\ref{model}.  Section~\ref{results} contains the quantitative
analysis of the circuit, and in particular of the role played by the
cellular state of growth on the dynamics. More specifically,
Subsection~\ref{symm-section} analyzes the symmetric repressilator,
showing the possibility of a growth-mediated switch between
oscillations and convergence to a stable fixed
point. Subsection~\ref{asymm-section} focuses on the asymmetric case,
with particular emphasis on the role of gene chromosomal position on
the circuit behavior. Finally, the last section discusses the
implications of the results from both a biological and a physical
modeling standpoint.

\section{Model}
\label{model}
\subsection{Growth-rate dependence of gene expression}
\label{growth-rate-dep}

This section reviews the approach of Klumpp and
coworkers~\cite{Klumpp2009}, and the necessary assumptions in order to
extend it to the analysis of circuit dynamics.  This modeling strategy
constitutes the basis for our description of the repressilator,
introduced in section~\ref{repressilator_intro}.

In absence of regulation, the dynamics of messenger RNA (mRNA) levels 
and protein concentrations, denoted with $m$ and $p$ respectively,
is described by two equations
\begin{eqnarray}
 \dot{m} ~&=&~ \frac{g ~\alpha_{m}} {V} -\beta_{m} m \nonumber\\
\dot{p}~&=&~\alpha_p m -\beta_p p \ ,
\label{basic}
\end{eqnarray}
where $\alpha_{m}$ and $\alpha_p$ are the transcription and
translation rates respectively, $\beta_{m}$ and $\beta_p$ the
degradation rates of mRNAs and proteins, and $V$ the cell volume. The
parameter $g$ is the gene copy number. If the gene \rev{support} is a
plasmid, $g$ represents the (mean) plasmid copy number. For genes
integrated on the chromosome, the copy number can vary because of DNA
replication, especially at fast growth rates, when multiple copies of
the genome are replicated at the same time.
This phenomenon whereby gene dosage is modified by DNA replication is
described by the Cooper-Helmstetter model~\cite{Cooper1968}, and
reviewed in Appendix~\ref{CH-model}.  Note that for fast
\textit{E.~coli} growth, $g$ increases with decreasing distance $\ell$
from the replication origin (illustrated in Fig.~\ref{fig1}B), since
the cell engages overlapping rounds of DNA replication in order to
allow fast growth~\cite{Grant2011}.

In bacteria, proteins are typically stable, with a lifetime longer
than the cell cycle, while mRNAs have a lifetime of just a few
minutes~\cite{Taniguchi2010}.  Therefore, the loss of protein is
mainly due to dilution through growth and cell division, so that an
effective degradation rate $\beta_p = \mu~ln2$ (where $\mu$ is the
growth rate) can be safely used in most cases.  On the other hand, the
fast time-scale of mRNA dynamics allows a quasi-equilibrium
approximation. Thus, Eq.~\ref{basic} can be reduced to
\begin{equation}
 \dot{p}~=~\frac{g~\alpha_{m} \alpha_p}{\beta_{m} V} -\beta_p p.
\end{equation}
Rescaling time with the dilution rate $\beta_p$, all the growth-rate
dependence can be factorized in a single term $F(\mu)$, 
\begin{equation}
  \dot{p}~=~\frac{g~\alpha_{m} \alpha_p}{\beta_{m} \beta_{p} V} - p =
  p^{*} F(\mu) -p \ . 
\label{constitutive}
\end{equation}

Normalizing the growth-rate dependence $F(\mu)$ such that $F(\mu)=1$
for $\mu= 1~db/hr$ (i.e. doublings per hour), the parameter $p^{*}$
represents the steady-state concentration of the \rev{constitutive
  (unregulated) gene} at $\mu=1~db/hr$. Thus, it is a measure of the
degree of basal expression.  \rev{Klumpp and
  coworkers~\cite{Klumpp2009}, refer to this quantity as the}
``promoter strength''. \rev{However,} this terminology might be
slightly misleading, as the term actually includes non-transcriptional
parameters, such as the translation efficiency or the gene copy number
at $\mu=1~db/hr$. \rev{Thus, we will refer to it as ``basal
  expression'' level in the following}.

\rev{In principle, all the cellular parameters in $F(\mu)$ may display
  a growth-rate dependence.  The volume $V$ is known experimentally to
  change with growth rate~\cite{Schaechter1958,Bremer1996}, with
  faster-growing bacteria being larger than slower-growing ones.  The
  protein degradation rate $\beta_p$ is a direct consequence of
  dilution due to cell growth and division, at least for stable
  proteins, and therefore, as previously discussed, it is a linear
  function of the growth rate.  By contrast, the empirical growth-rate
  independence of the mRNA degradation rate $\beta_m$ has a less
  obvious interpretation~\cite{Klumpp2009}.
  In order to sustain fast growth and a large cell mass, the cellular
  abundance of the transcriptional and translational machinery is
  expected to increase at fast growth, and indeed this is what can be
  observed experimentally~\cite{Bremer1996}.  However, the rates of
  transcription and translation ($\alpha_m$ and $\alpha_p$) of a gene
  only reflect the availability of ``free'' (active but not already
  engaged in any process) RNA polymerases and ribosomes. This fraction
  can be quite hard to estimate~\cite{Klumpp2008}, requiring several
  empirical measurements, and is in general not merely proportional to
  the total abundance.  In fact, while the cellular ribosome content
  increases strongly with growth rate, the translation rate is
  apparently constant in different nutrient
  conditions~\cite{Klumpp2009}.  The gene dosage $g$ is probably the
  best-characterized parameter, since it is captured by the
  Cooper-Helmstetter model~\cite{Cooper1968}. Its increase with growth
  rate is a consequence of the coupling between DNA replication and
  growth.  To sum up, while it is possible to try to rationalize at
  least some of the growth-rate dependences of the basic parameters,
  there is in general no \emph{a priori} quantitative expectation (and
  sometimes not even a basic intuition) for them, and thus for
  $F(\mu)$.  Hence, $F(\mu)$ can be fully defined only through the
  empirical knowledge of the growth-rate dependences of the different
  parameters.  }

For \textit{E.~coli}, the numerical values of these parameters were
collected for five growth rates $\mu$ between $0.6~db/hr$ and
$2.5~db/hr$\secondrev{~\cite{Klumpp2009}. 
These values are reported in the supplementary material of Klumpp \textit{et al.} (Cell 2009, reference~\cite{Klumpp2009}), more specifically in Table S1. 
The growth-rate dependences of the various parameters are plotted in Figure 1 of the same paper.}
\secondrev{As a combination of these parameters}, $F(\mu)$ decreases in 
a weakly nonlinear fashion in the available range \secondrev{of growth rates} (interpolation will
be used in the following when needed). 
\secondrev{The growth-rate dependence of $F(\mu)$, and thus of the concentration of the protein product of 
a constitutive gene, is reported in Figure 2D and in Figure 3 of the same reference~\cite{Klumpp2009}.}

The empirical growth-rate dependences that define $F(\mu)$ are based
on experimental measurements of the average cellular properties in a
growing cell population. Thus, they are in principle also dependent on
the age distribution (where ``age'' stands for stage of the cell
cycle) of the population. However, the differences between averages
calculated over a cell cycle and over a cell population with the age
distribution determined by the exponential growth are not
quantitatively significant, as discussed in detail in
Appendix~\ref{app1}.  \rev{Therefore, the available empirical
  measurements of all the key parameters, which are based on
  population averages, will be used throughout this paper as an
  approximation of their cell-cycle averages.}

Moreover, we will not consider explicitly the
cell-cycle dynamics of cellular parameters, which, for sufficiently
long time scales, can be averaged out using for example effective gene
dosage or effective protein dilution rate. As a consequence,
Eq.~\ref{constitutive} can accurately describe the dynamics of protein
concentration on time scales longer than the generation
time~\cite{Marathe2012}, as it will always be the case in the
following.

\subsection{The repressilator}
\label{repressilator_intro}

\begin{figure}
\begin{center}
\epsfig{file=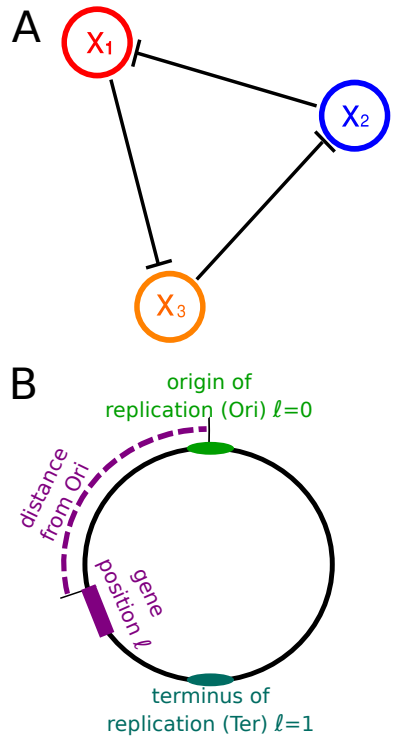, width=5cm}\\
\caption{(Color online) \textbf{A. Scheme of the repressilator.} Each
  node represents the expression level of one gene, links represent
  repressive regulatory interactions, as described by
  Eq.~(\ref{eq:repr}).\textbf{ B. Scheme of the \emph{E.~coli}
    chromosome}, illustrating the coordinate $\ell$, used for the
  positions where a gene can be inserted. \label{fig1}}
\end{center}
\end{figure}

\rev{The repressilator is a genetic network consisting of three genes,
  each of which encodes for a ``transcription factor'' (i.e. a protein
  able to control the transcription level of one or a set of target
  genes by binding to their upstream genomic region) that represses
  the expression of its target in a cycle of regulations
  (Fig.~\ref{fig1}A).  The effect of this negative regulation can be
  modeled phenomenologically as a multiplicative factor rescaling the
  target production rate (the positive term on the right-hand side of
  Eq.~\ref{constitutive}) with a non-linear function of the repressor
  concentration (called ``Hill function'')~\cite{Bintu2005,Alon2006}:}
\begin{equation}
R(x/k)=\frac{1}{1+(x/k)^n} \ ,
\label{eq:repr}
\end{equation}

where the Hill coefficient $n$ defines the degree of cooperativity
(determining the steepness of $R$), while the dissociation constant
$k$ specifies the repressor concentration at which the production rate
is half of its constitutive value.  \rev{ With this mathematical
  representation of transcriptional regulation, } the repressilator
circuit can be described by three equations, one for each gene, based
on the gene expression model discussed \rev{in the previous section}:
\begin{eqnarray}
\dot{x_1}~&=&~x_{1}^{*} F_{1}(\mu) R(x_{2}/k_1) -x_1\nonumber\\
\dot{x_2}~&=&~x_{2}^{*} F_{2}(\mu) R(x_{3}/k_2) -x_2\nonumber\\
\dot{x_3}~&=&~x_{3}^{*} F_{3}(\mu) R(x_{1}/k_3) -x_3 \ ,
\end{eqnarray}
where the terms $x^{*}_{i}$ are the basal expression levels\secondrev{. They 
represent the steady-state concentration of protein $i$ in 
absence of repression at $\mu=1~db/hr$. Therefore,  the ``basal expression level'', as defined here for a negatively regulated gene, is  
the maximal expression level that can be achieved in a fixed growth condition ($\mu=1~db/hr$) 
in absence of regulation~\footnote{\secondrev{Note that the term ``basal expression'' is sometimes found in the
literature with a different meaning, i.e. the low but detectable level
of expression that a negatively regulated gene maintains in conditions
of strong repression (also called ``leakage'' level of expression).}}.}

 The growth rate functions 
$F_{i}(\mu)$ can be gene-specific due to the dependence of gene 
dosage $g_i$ on the chromosomal gene position. 

Assuming similar dissociation constants for the three promoters
$k_1\simeq k_2 \simeq k_3 =k$, the protein concentrations can be
rescaled with $k$
\begin{eqnarray}
\dot{x_1}~&=&~x_{1}^{*} F_1(\mu) R(x_{2}) -x_1\nonumber\\
\dot{x_2}~&=&~x_{2}^{*} F_2(\mu) R(x_{3}) -x_2\nonumber\\
\dot{x_3}~&=&~x_{3}^{*} F_{3}(\mu) R(x_{1}) -x_3.
\label{repress_general}
\end{eqnarray}

With these notations, the \rev{basal expression levels} $x_{i}^{*}$ have
dimensionless units, as they are protein concentrations rescaled with
the dissociation constant $k$.

\begin{figure*}[ht!!]
\begin{center}
\epsfig{file=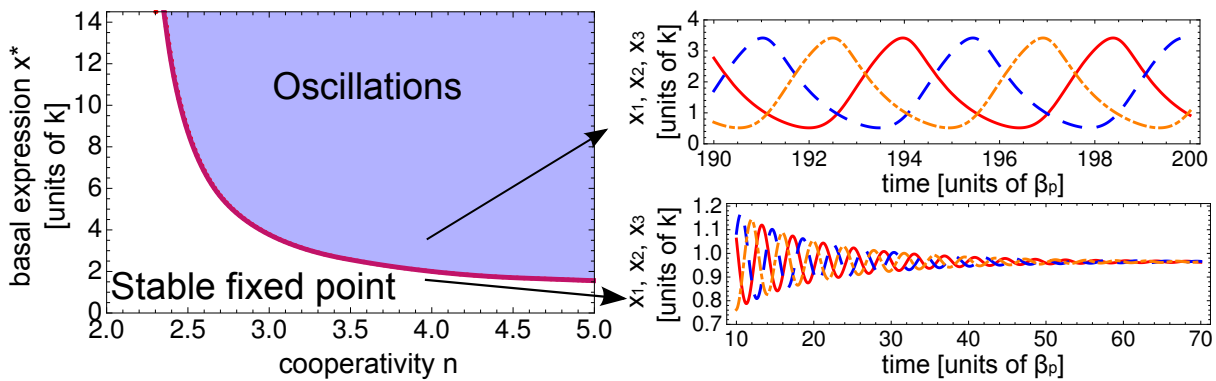, width=16cm}\\
\caption{(Color online) \textbf{Stability diagram of the symmetric
    repressilator.} The left panel shows the parameter space giving
  rise to oscillations. The \rev{basal expression} $x^*$ is
  dimensionless, measured as protein concentration at $\mu=1~db/hr$ in
  absence of repression in units of the dissociation constant
  $k$. High repression cooperativity and \rev{high basal expression levels
    (``strong promoters'')} can ensure stable oscillation.  The right
  panel plots are illustrative examples of the dynamics in
  correspondence with two sets of parameter values
  ($\{n=4,x^{*}=1.8\}$ and $\{n=4,x^{*}=5\}$), showing the time
  evolution of the three protein concentrations $x_1,x_2,x_3$ (in
  units of $k$). The red continuous curve describes the dynamics of
  $x_1$, the blue dashed curve $x_2$, and the orange dot-dashed curve
  $x_3$.
  \label{fig2}}
\end{center}
\end{figure*}

\section{Results}
\label{results}

\subsection{Growth rate affects qualitatively and quantitatively the
  dynamics of a symmetric repressilator}
\label{symm-section}
 
We start the analysis from the simplified case of a symmetric
repressilator, i.e. the three genes have approximately the same
production/degradation rates for mRNA and proteins, as well as the
same growth-rate dependence of parameters (this is attained for
example when all genes are integrated on a single plasmid or at a
similar distance from the origin of replication on a chromosome). This
is the most studied case, and was approached with different
mathematical
descriptions~\cite{Purcell2010,Mueller2006,Strelkowa2010,Buse2010}. The
symmetric approximation was originally proposed to explain the
behavior of the synthetic realization \textit{in
  vivo}~\cite{Elowitz2000}.

In the symmetric case, the functions encoding the growth-rate
dependence $F(\mu)$ and the \rev{basal expression levels} $x^*$ are exactly the
same for each of the three genes, simplifying
Eq.~\ref{repress_general} to
\begin{equation}
 \dot{x_i}~=~x^{*} F(\mu) R(x_{i+1}) -x_i \ ,
\label{simm_rep}
\end{equation}
where $i\in[1,3]$ and $x_4 \equiv x_1$.

\subsubsection{At fixed growth rate, the dynamics is determined by
  \rev{basal expression} and cooperativity}

Before addressing the effects of the growth-rate dependence of
parameters, we characterize the circuit dynamics at fixed growth
conditions, for simplicity at $\mu=1~db/hr$ where $F(\mu)=1$.  In
general, the symmetric repressilator can display stable oscillations,
arising through a Hopf bifurcation~\cite{Mueller2006}. More
specifically, Eqs.~\ref{simm_rep} have an oscillatory solution if the
condition for cooperativity $n>2$ is satisfied, as can be shown in a
straightforward way considering the symmetry of the
system~\cite{Blossey2008}.

For a given steepness of the repression function satisfying the
condition $n>2$, the stability of the limit cycle is solely determined
by the \rev{basal expression} $x^*$. The values of $x^*$ that ensure
stability of the oscillatory state can be calculated using linear
stability analysis and the resulting stability diagram is shown in
Fig.~\ref{fig2}. Essentially, an increase of either cooperativity or
\rev{basal expression} can help the stabilization of oscillations. This
result is common to all the different repressilator descriptions
proposed in the literature~\cite{Purcell2010}.

\begin{figure}[ht!]
\begin{center}
\epsfig{file=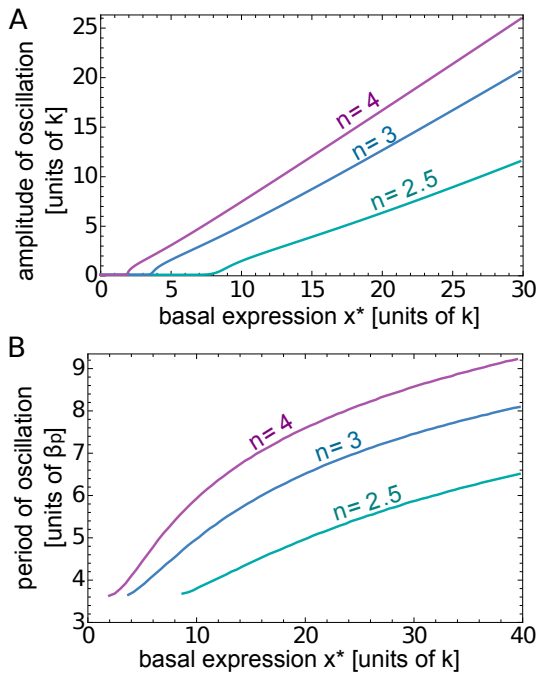, width=7cm}\\
\caption{(Color online) \textbf{Amplitude and period of oscillation of
    a symmetric repressilator.}  The plots show oscillation amplitude
  (A) and period (B) as a function of the \rev{basal expression level}
  for three different values of cooperativity $n$. Both quantities are
  plotted in rescaled (dimensionless) units. The amplitude
  (concentration) is rescaled by the dissociation constant $k$, while
  the period has units of protein degradation rate $\beta_p$. For each
  value of $n$, Eqs.~\ref{simm_rep} were integrated numerically with
  values of $x^*$ in the range $0.1-40$ and step-size $0.1$. For each
  numerical solution, the oscillation amplitude and period were
  evaluated after convergence to a stable limit cycle.
  \label{fig3}}
\end{center}
\end{figure}

Furthermore, the two parameters $n$ and $x^*$ determine the amplitude
and period of oscillation (Fig.~\ref{fig3}).  The approximately linear
and logarithmic dependences of amplitude and period respectively that
emerge from numerical integration of Eqs.~\ref{simm_rep} can be
rationalized by the following rough but conceptually simple
argument. Each gene in the repressilator tends to oscillate between
two states corresponding to its maximally repressed state and its
maximally activated state. In a simplified situation of a step
repression function ($n \gg1$) that switches the target gene on and
off after its equilibration, the protein concentration would go from
$x_i\simeq 0$ (fully repressed) to $x_i\simeq x^*$ (fully activated),
thus leading to a linear dependence on \rev{basal expression} of the
oscillation amplitude. On the other hand, the oscillation period
depends on the timescales of gene activation and deactivation. For
example, the time $\tau$ required to go from the fully activated state
$x_i(t)\simeq x^*$ to $x_i(t)\simeq 1$, \rev{when, assuming a
  step-like repression function, it releases the repression of its
  target}, is given by $x^* e^{-\tau}=1 \rightarrow
\tau=ln(x^*)$. This expression suggests that the \rev{basal
  expression} contributes logarithmically to the deactivation
timescale, which is compatible with the dependence of the period on
$x^*$ measured by numerical integration shown in Fig.~\ref{fig3}B.

\subsubsection{Increasing the growth rate can destabilize oscillations}

\begin{figure}
\begin{center}
\epsfig{file=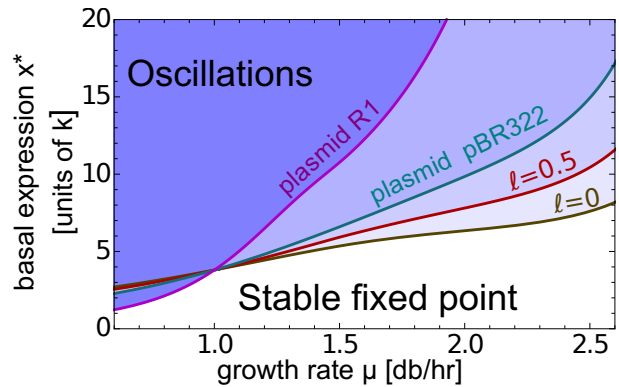, width=8cm}\\
\caption{(Color online) \textbf{The repressilator has different
    dynamics in different growth conditions.} At fixed repression
  cooperativity ($n=3$ in this plot), the repressilator can show
  stable oscillation or convergence to a steady state depending on the
  cell growth rate $\mu$ and the \rev{basal expression} $x^*$ (in
  units of $k$). The circuit \rev{physical support} (plasmid,
  chromosome) defines the range of growth rates at which a limit cycle
  can be observed. The cases of circuit integration on plasmids ($R1$
  and $pBR322$) and on chromosome near the origin of replication
  ($\ell=0$) and between the origin and the terminus ($\ell=0.5$) are
  represented in the plot. \label{fig4}}
\end{center}
\end{figure}

\rev{Let us now consider a specific experimental realization of the
  repressillator, which uses a set of genetic components with some
  intrinsic parameters.} Given the \rev{basal expression level} and
the cooperativity of repression, defined by the specific properties of
the \rev{actual} genes and promoters implementing the repressilator,
the growth-rate dependence of parameters defines a vertical path in
the stability diagram of Fig.~\ref{fig2}, since the ``effective"
\rev{basal expression} $F(\mu)~x^*$ decreases nonlinearly with growth
rate. This path can cross the border of stability of the limit cycle,
defining a maximum growth rate at which stable oscillations can be
sustained.  \rev{More precisely, the basal expression level $x^*$
  defines the position of the path at $\mu=1~db/hr$, while the
  function $F(\mu)$ is related to its length. In fact, given the
  experimentally accessible growth rates, $F(\mu)$ identifies the span
  of effective basal expressions that can be explored changing the
  growth rate, and thus the upper and lower bound of the path.
  Therefore}, both factors contribute to establish the range of growth
rates in which oscillations are expected \rev{experimentally}.

Remarkably, the circuit can show qualitatively different dynamics
depending on growth rate, and hence on nutrient conditions.
Fig.~\ref{fig4} shows this for the case of cooperativity
$n=3$. Sustained oscillations can be observed at slow growth, while
in fast-growth conditions the dynamics can converge to a stable fixed
point.  The parameter range where oscillations are stable depends on
where the repressilator is integrated through the gene dosage factor
$g$ appearing in $F(\mu)$ (see Eq.~\ref{constitutive}).

If the repressilator is integrated on a plasmid, as the original
\textit{in vivo} experimental realization~\cite{Elowitz2000}, the gene
dosage $g$ is simply given by the plasmid copy number, which has a
plasmid-specific growth-rate dependence that can be quite
strong~\cite{Klumpp2009,Klumpp2011}.  Fig.~\ref{fig4} compares the
parameter regions corresponding to convergence to stable oscillation 
or to a fixed point for a repressilator integrated on the two plasmids
$R1$ and $pBR322$ for which the copy number has been measured in
different growth media~\cite{Klumpp2009}.

The dosage of a chromosomal gene, instead, is determined by its
genomic position as set by the Cooper-Helmstetter model (see
Appendix~\ref{app1} and Fig.~\ref{fig1}B). Therefore, for a
repressilator integrated on the chromosome, the normalized circuit
distance $\ell$ from the replication origin (Ori) defines the growth-rate
dependence of the dynamics (where $\ell=0$ represents Ori and $\ell=1$ the
replication terminus, Ter, Fig~\ref{fig1}B). It should be noticed that
the three genes composing the repressilator are assumed here to be
inserted approximately in a single chromosomal location (or
equivalently in different replichores, the oppositely replicated
chromosome halves, but at the same distance $\ell$ from Ori, see
Fig.~\ref{fig1}B).  The effects of the imbalance in gene dosage
generated by different gene locations on the chromosome are explored
in section~\ref{asymmetry_growth}.

\subsubsection{Period and amplitude of oscillation are growth rate
  dependent}

\begin{figure}
\begin{center}
\epsfig{file=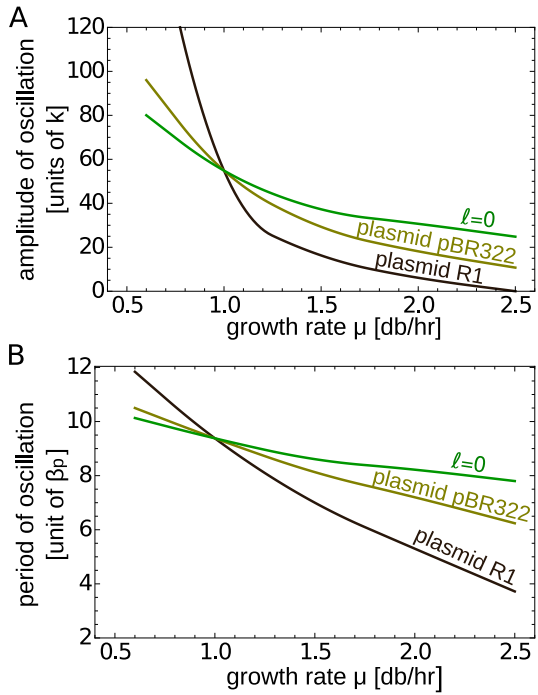, width=7cm}\\
\caption{(Color online) \textbf{The oscillation amplitude and period
    of a symmetric repressilator vary significantly with growth rate.}
  The plots show the scaling with growth rate of oscillation amplitude
  (concentration in units of $k$) and period (time in units of protein
  degradation rate) for a ``strong promoter'' (\rev{basal expression}
  $x^*=70$ in $k$ units) and cooperativity $n=3$. The three curves
  correspond to integration of the repressilator on plasmid $R1$,
  plasmid $pBR322$, and to chromosomal integration of the three genes
  near the origin of replication ($\ell=0$). A repressilator located
  near the replication terminus ($\ell=1$, not shown) follows a
  growth-rate dependence of amplitude and period similar to the case
  of plasmid $pBR322$.
  \label{fig5}}
\end{center}
\end{figure}

Fig.~\ref{fig4} shows that increasing the \rev{basal expression} $x^*$
(which we \rev{recall} also includes parameters not based on transcription)
and the steepness of the repression function $n$ can make the limit
cycle solution stable in a wide range of conditions.
However, even if a repressilator is designed to exhibit oscillations
in the experimentally accessible conditions, the growth rate is still
expected to influence the oscillation amplitude and period in a
measurable way (i.e. by a factor that can exceed five,
Fig.~\ref{fig5}).
Indeed, the effective \rev{basal expression}
$F(\mu)~x^*$ decreases with increasing growth rate, because of the
functional form of $F(\mu)$, leading to a reduced amplitude and period
of oscillation.  Integrating Eqs.~\ref{simm_rep} numerically for
parameter values corresponding to different growth rates allows to
predict the oscillation period and amplitude for a chromosomally
integrated repressilator or for a plasmid implementation, if the
scaling of the plasmid copy number with growth rate is known, The
presence of nonlinearities in the system makes the dependence on
growth rate of these two variables nontrivial, as represented in
Fig.~\ref{fig5}.

\subsection{The effect of intrinsic and position-induced asymmetry on
  the repressilator dynamics}
\label{asymm-section}

\begin{figure*}
\begin{center}
\epsfig{file=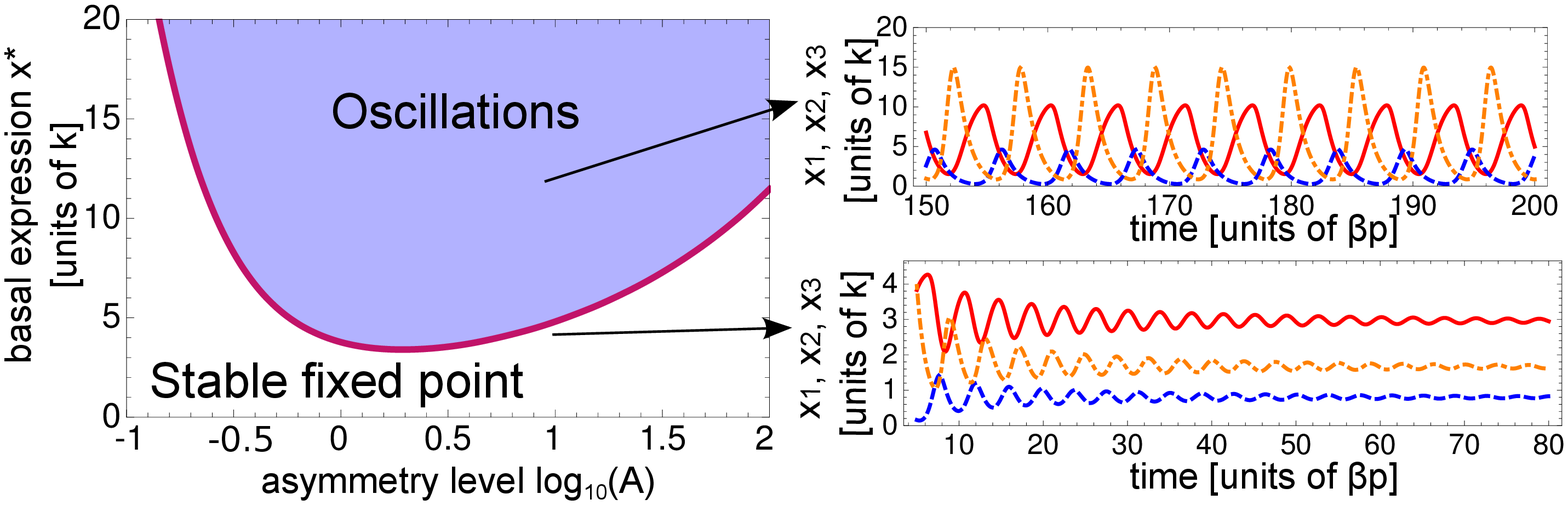, width=16cm}\\
\caption{(Color online) \textbf{Stability diagram of the asymmetric
    repressilator} The parameter space is divided in the regions
  corresponding to convergence of the dynamics to a limit cycle and to
  a stable fixed point. The two parameters considered are the
  \rev{basal expression} $x^*$ (in units of $k$) and the level of
  asymmetry $A$ (the factor by which the \rev{basal expression} of one
  of the genes differs from the two others). The cooperativity is
  fixed to $n=3$ but the qualitative shape of the phase diagram holds
  for cooperativity values $n>2$. While a high degree of asymmetry
  reduces the robustness of the oscillatory state, a slight asymmetry
  can stabilize the damped oscillations showed by the symmetric
  system. The right panel shows an example of the dynamics of the
  three protein concentrations for two sets of parameter values
  ($\{x^*=12,A=10\}$ and $\{x^*=4.5,A=10\}$). The red continuous curve
  describes the dynamics of $x_1$, the blue dashed curve refers to
  $x_2$, and the orange dot-dashed to $x_3$.\label{fig6}}
\end{center}
\end{figure*}

Differences in intrinsic properties of the genes composing the
circuit, such as affinity for RNA polymerase or ribosome binding, or
gene dosage imbalances due to the specific gene location on the
chromosome, lead to unequal parameter values in the equations for the
dynamics of $x_i$.
This situation is generally referred to as ``asymmetric''.  A certain
degree of asymmetry is expected for generic synthetic realizations of
the repressilator. However, with the exception of a few
studies~\cite{Samad2005,Strelkowa2010}, its consequences on the
dynamics have \rev{not been fully characterized}
theoretically~\cite{Purcell2010}.  In the modeling framework adopted
here, intrinsic gene properties are \rev{summarized} by the \rev{basal
  expression level}. Additionally, using the Cooper-Helmstetter model (see
Appendix~\ref{app1}) it is possible to account for the
position-dependent scaling of gene dosage with growth rate.

The two possible contributions to repressilator asymmetry, intrinsic
gene properties and gene dosage, and their effects on the dynamics can
be analyzed separately.
We first address the dynamics at fixed growth rate of a repressilator
composed of genes with different \rev{basal expression levels}, and subsequently
explore the consequences of position-induced asymmetry at different
growth rates for a repressilator made of genes with equal intrinsic
properties, but different chromosomal location.

\subsubsection{Effects of asymmetry at fixed growth rate}

We consider the simplified situation in which only one of the genes of
the repressilator differs from the others in its \rev{basal expression} by
a factor $A$, but the three genes share the same growth rate
dependence $F(\mu)$. Thus, the single additional parameter $A$
introduced in the model measures the level of circuit asymmetry.  The
system of equations describing the dynamics of a repressilator
designed this way is
\begin{eqnarray}
\dot{x_1}~&=&~x^{*} F(\mu)  R(x_{2}) -x_1\nonumber\\
\dot{x_2}~&=&~ x^{*} F(\mu) R(x_{3}) -x_2\nonumber\\
\dot{x_3}~&=&~A~ x^{*} F(\mu) R(x_{1}) -x_3 \  .
\label{asym_eq}
\end{eqnarray}

At fixed growth rate (for simplicity we take the case of growth rate
$\mu=1 db/hr$, where $F(\mu)=1$), linear stability analysis can be
applied to study the fixed point stability. As in the symmetric case
described above, a Hopf bifurcation stands between the system
convergence to a stable fixed point and the oscillatory solution. For
each repression cooperativity level $n$, the stability diagram can be
drawn as a function of the \rev{basal expression} $x^*$ and the
asymmetry level $A$, as shown in Fig.~\ref{fig6} for $n=3$.  This
diagram essentially shows that a high degree of asymmetry destabilizes
the oscillations. Therefore, if the goal is to engineer a stable
oscillator, a roughly symmetric design is generally preferable.

However, the minimum of the boundary curve between the two asymptotic
dynamical behaviors (\rev{continuous} purple curve in Fig.~\ref{fig6}) does not
correspond to the symmetric case ($log_{10}(A)=0$).  This result indicates
that a symmetric system showing damped oscillations (as it is
generally the case for parameter values just below the boundary in
Fig.~\ref{fig6}) can be pushed toward a stable oscillation regime by
slightly increasing the \rev{basal expression level} of just one gene, if the
resulting asymmetry is not too strong.

\begin{figure}
\begin{center}
  \epsfig{file=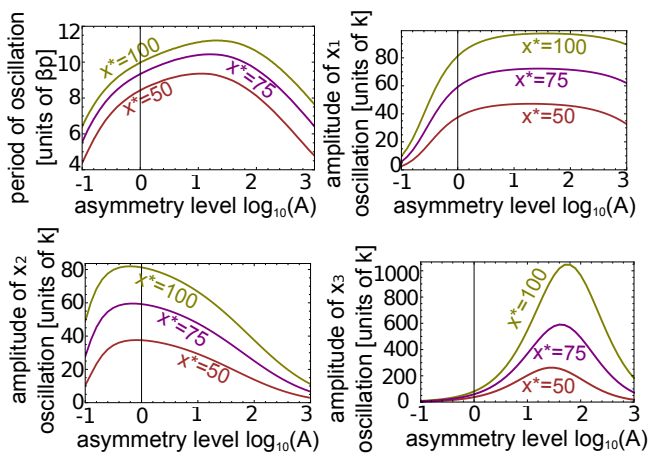, width=8.5cm}\\
  \caption{(Color online) \textbf{Influence of asymmetry on period and
      amplitude of oscillation.} The figure shows the nonlinear
    effects of the asymmetry induced by the presence of gene $x_3$
    with a \rev{basal expression level} that differs by a factor $A$
    from the other genes' \rev{basal expression} $x^*$. The period
    (upper left plot) and the gene specific oscillation amplitudes
    (other plots) are shown as a function of the asymmetry level
    $log_{10}(A)$ for three \rev{basal expression} values that ensure
    oscillations in the whole range of asymmetry explored.  The curves
    are obtained by measuring period and amplitude of numerical
    solutions of Eq.~\ref{asym_eq} for values of $log_{10}(A)$ spaced
    by $0.01$ in the range $\{-1,3\}$.
    \label{fig7}}
\end{center}
\end{figure}

Moreover, the presence of a gene with a different \rev{basal expression}
breaks the symmetry in the dynamics of protein concentrations, making
the oscillation amplitude (or the stable fixed point) gene-specific,
as shown in Fig.~\ref{fig6} for two parameter sets. 
Thus, a \rev{possible} test of the effective symmetry of experimental
repressilator realizations would entail measuring the oscillation
amplitude of two fluorescently-tagged protein products of genes in the
circuit.  More specifically, the level of asymmetry, introduced by the
presence of a gene with different intrinsic properties, affects the
oscillation period and amplitude in a predictable way and with a
gene-specific signature on the oscillation amplitude
(Fig.~\ref{fig7}).  Therefore, the global dynamical properties of the
repressilator can be tuned simply by changing the parameters relative
to a single gene.

\subsubsection{The chromosomal position of genes affects the circuit
  dynamical properties at varying growth rates}
\label{asymmetry_growth}

\begin{figure*}
\begin{center}
\epsfig{file=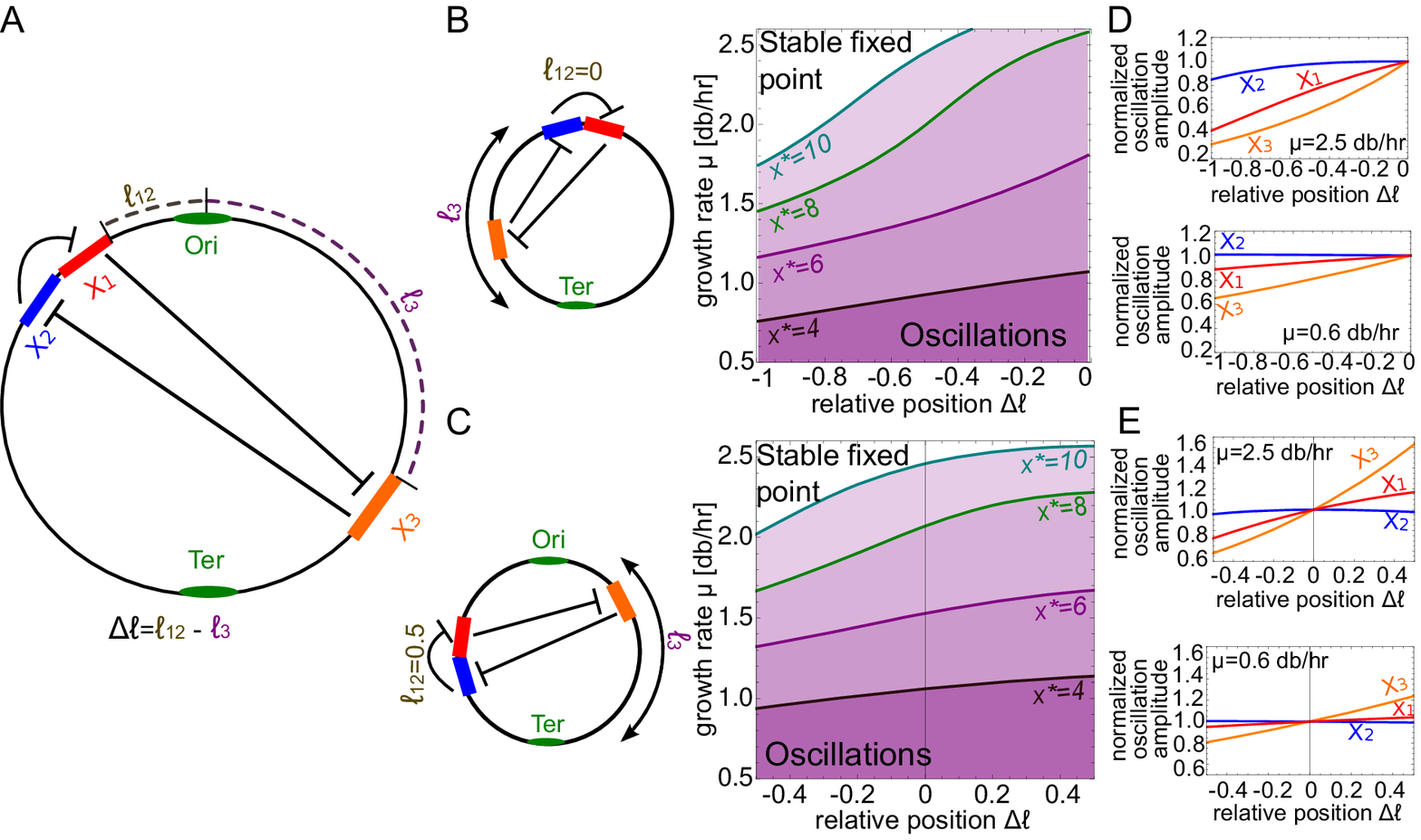, width=17cm}
\caption{(Color online) \textbf{Effect of gene-dosage induced
    asymmetry on oscillation stability} (A) Scheme of the
  repressilator chromosomal configuration analyzed here.  Two genes
  are placed at approximately the same distance $\ell_{12}$ from the
  origin of replication (possibly on opposite arms), while the third
  is in a different position $\ell_3$. This genomic configuration
  induces a growth-rate dependent copy-number imbalance between genes,
  as a consequence of DNA replication.  The difference $\Delta \ell =
  \ell_{12}-\ell_3$ measures the effective asymmetry level ($\Delta
  \ell=0$ corresponds to the symmetric case) since the ratio between
  the gene dosages (the factor $A$ in Eq.~\ref{asym_eq}) is $2^{\mu C
    \Delta \ell}$, where $C$ is the DNA replication time. (B,C) The
  region of oscillation stability is plotted as a function of growth
  rate $\mu$ and relative position $\Delta \ell$ for different values
  of the \rev{basal expression} $x^*$. The genomic configurations
  analyzed, and thus the values of $\Delta \ell$, are obtained moving
  one gene along the chromosome while keeping the others near the
  replication origin ($\ell_{12}=0$) in plot B, or in a mid position
  ($\ell_{12}=0.5$) in plot C (see the corresponding schemes on the
  left).  The plots show how gene positions influence the range of
  growth rates in which stable oscillations take place.  (D,E) Effect
  of the relative position $\Delta \ell $ on the oscillation
  amplitude, normalized with its value in the symmetric case, of the
  three proteins is plotted for two different growth rates (\rev{the
    basal expression} is $x^* =75$ in units of $k$). The sets of
  positions considered in D and E are those shown in plot B and C
  respectively.  The change in oscillation amplitude that can be
  obtained by placing the same genes in different configurations
  depends on the growth conditions. \label{fig8}}
\end{center}
\end{figure*}

To isolate the effect of genomic position of genes on the dynamics of
a chromosomally integrated repressilator, we consider the case of a
circuit that is symmetric in terms of intrinsic gene properties, but
where a gene is placed at a varying distance from the replication
origin. This situation is illustrated in Fig.~\ref{fig8}A, where the
genes with protein product concentrations $x_1$ and $x_2$ are placed
in contiguous positions, thus at approximately the same distance
$\ell_{12}$ from Ori, while the gene corresponding to $x_3$ is in
position $\ell_3$.
In modeling terms, the equivalence of intrinsic gene properties
corresponds to the assumption of identical effective \rev{basal
  expression} $ x_i F_i(\mu)$ at extremely slow growth, where the
growth-dependence of gene dosage $g_i$ for each gene $i$ is
irrelevant, since $g_i\simeq 1$ for every genomic position.  In other
words, the following relation is satisfied: $x^* ~F(\mu \simeq
0)~=~x_3^* ~F_3(\mu \simeq 0)$, where $x^*$ and $F(\mu)$ are the
\rev{basal expression level} and growth-dependence function of genes
$1$ and $2$, and $F_3(\mu)$ differs from $F(\mu)$ only because of the
position-dependent scaling of gene dosage with growth rate.

It is straightforward to verify that the dynamics of a repressilator
satisfying the above conditions is described by Eq.~\ref{asym_eq} with
$A= 2^{\mu C \Delta \ell}$, where $\Delta \ell= \ell_{12} -\ell_3$ and $C$ is the
time required for DNA replication ($C\simeq 40~min$ for fast
growth~\cite{Michelsen2003}). The factor $A$ is simply the ratio
between the gene copy numbers, $g_3/g_{1(2)}$ as given by the Cooper-
Helmstetter model (Appendix~\ref{app1}), and encodes the growth-rate
dependent level of asymmetry induced by gene position in an otherwise
symmetrical circuit.
Note that the sign of the relative position $\Delta \ell$ has relevance.
For example, the configuration with two genes near the replication
origin and the third near the terminus ($\Delta \ell=-1$) leads to a
quantitatively different dynamics with respect to the opposite
configuration with $\Delta \ell=1$. The repressilator dynamics can be
analyzed for different values of $\Delta \ell$ to explore the effects of
gene position.

Fig.~\ref{fig8}B shows a stability diagram obtained moving one gene
along the chromosome while the other two genes are near the
replication origin ($\ell_{12}=0$), for different levels of \rev{basal
  expression} $x^*$. Analogously, the stability diagram in Fig.~\ref{fig8}C
is obtained keeping $\ell_{12}=0.5$ and varying the third gene
position $\ell_{3}$.  Interestingly, this analysis shows that at fixed
growth rate the repressilator can converge to a limit cycle or to
oscillations depending on where genes are integreted on the
chromosome, highlighting the importance of including the chromosomal
gene coordinates as variables in genetic circuit models. The
positional effect is even more evident if two gene positions are
varied (data not shown).

As discussed in the previous section, the oscillation stability can be
reinforced by increasing the \rev{basal expression} of one gene. A simple
way to increase the  \rev{basal expression} of one gene for a
chromosomally integrated circuit is moving its position toward Ori,
since this increases its average gene copy number. Indeed, gene
position has been used to modulate the oscillation features in an
experimental synthetic implementation of another genetic
circuit~\cite{Atkinson2003}. However, as we show for our case, the
effect of the displacement of a gene is growth-rate dependent. In
fact, while moving a gene from Ter to Ori allows an increment of its
gene dosage of a factor $4$ for a $\mu=3~db/hr$, this factor
decreases with doubling time up to $1$ for slow growth.  The
consequence of this observation on the repressilator dynamics is
evident looking at Fig.~\ref{fig8}B and C. The same change in configuration
(change in $\Delta \ell$) has different effects depending on the growth
rate.  
The modeling framework adopted here gives a quantitative prediction
of the gain in oscillation stability that can be achieved by moving
the genes along the chromosome depending on the experimental growth
conditions, giving an experimental guideline for the best gene
insertion sites in order to obtain the desired dynamics.
Finally, Fig.~\ref{fig8}D and E show the oscillation amplitude dependence 
on the relative gene positions for the configurations 
described in Fig.~\ref{fig8}B and C respectively. This dependence changes  
significantly at different growth rates,  
showing that the effect of gene position is strongly 
influenced by the cellular environment, and thus by growth rate.  
This feature is also relevant for synthetic biology, since identical 
experiments carried out with different nutrient levels could in
principle lead to different results.

\section{Discussion}
\label{discussion}

To sum up, this work addresses the dynamics of a paradigmatic
bacterial oscillatory gene circuit, the repressilator, using for the
first time a modeling framework that accounts for relevant
physiological parameters, through \rev{their growth-rate
dependence}~\cite{Klumpp2009}.
From the modeling viewpoint, this framework entails assuming that the
parameters of the dynamical system are in fact dependent on a hidden
``super-parameter'', the growth rate, which \rev{encapsulates} cell
physiology.  Additional models or experiments are required in order to
obtain the dependence of each relevant parameter from the
super-parameter. For the case of \textit{E.~coli}, all this
information is available. \rev{The parameters change their value
  following the hidden variable, which makes the phenomenology of the
  dynamical system with respect to the standard parameters less
  informative. Specifically, ignorance of the role of the
  super-parameter and its behavior prevents from obtaining the
  physically relevant phase diagram.}  

%
\emph{A growth-rate induced ``dynamical switch''.}  Our results show
that the dynamics is dependent on the growth rate in a fashion that is
both qualitative and quantitative. Specifically, a symmetric
repressilator will lose its oscillatory state with increasing growth
rate unless its \rev{basal expression} is sufficiently high, and this
phenomenology is expected to be observable in a wide range of
experimentally accessible conditions.
Indeed, the growth-mediated switch between different dynamics shown in
Fig.~\ref{fig4} should be simple to observe experimentally, given the
typical values of the parameters involved. The average protein copy
numbers per cell span different orders of magnitude, from $10^{-1}$ to
$10^4$~\cite{Taniguchi2010}, while the values for the dissociation
constants have been reported to range between a few
molecules~\cite{Bintu2005a} and a few thousands~\cite{Setty2003}.
With these numbers, the \rev{basal expression} values (i.e. protein
concentration at $\mu=1$ in units of $k$) analyzed here, such as
$x^*\in(0,20)$ in the example in Fig.~\ref{fig4}, are well in the
physiological range. This suggests that a repressilator dynamics
characterized by loss of the oscillatory behavior at a critical growth
rate should \rev{not be too difficult to observe in the laboratory},
and that changes in both oscillation period and amplitude \rev{should
  also be quite likely} to be measurable experimentally.
\footnote{Note however that some of the dissociation constant values
  reported in the literature were measured \textit{in vitro} and thus
  could be underestimated with respect to \textit{in vivo} values,
  where non-specific binding plays an important role.}

\emph{Dependence of the dynamics \rev{on the physical support}. }
Furthermore, the growth-rate dependent behavior of the circuit is
\rev{affected by the support} it is embedded in, a plasmid or a
chromosome, and for a chromosome on the detailed coordinates of the
three promoters.
Fig.~\ref{fig4} and~\ref{fig5} suggest \rev{that} the stability of
oscillatory behavior is more sensitive to variations in growth rate
for a repressilator integrated on plasmids than on the chromosome, as
plasmid copy number can be strongly growth-rate dependent (as well as
variable from cell to cell).
However, integration on a high copy-number plasmid can naturally
increase the protein concentration (hence the ``promoter strength''
parameter defined \rev{by Klumpp and coworkers}), thus leading to more
robust oscillations.
Therefore, if the goal is engineering a stable synthetic oscillator,
there is probably a trade-off between the advantage of an increased
\rev{basal expression} typical of a high copy number plasmid, and the
unavoidable plasmid-specific growth-rate dependence (and cell-to-cell
variability) of the copy number~\cite{Tal2012}. 

On the other hand, for chromosomally integrated repressilators, the
dynamics depends nontrivially on gene position.
Recently, it has been shown that the spatial ordering of a set of
``important'' genes along the chromosome is strongly conserved between
different bacterial species and largely corresponds to their
expression pattern during growth~\cite{Sobetzko2012}, pointing to a
functional role for gene chromosomal position. 
The example of a putative chromosome-integrated repressilator analyzed
here suggests that the dynamics of genetic networks, in fast-growing
bacteria, should be influenced by the genomic position of its
components.  For example, for the repressilator case
(Fig.~\ref{fig8}), a variation in the relative position of genes
involved in a regulatory circuit can have different consequences
depending on the growth rate.  In this perspective, the analysis of
the phenotypic consequences of chromosomal rearrangements, such as
large inversions~\cite{Esnault2007}, should be revisited taking into
account the growth conditions.
More generally, it is tempting to speculate that the evolutionary
pressure for keeping a certain gene order with respect to genome
replication may be partially due to natural selection of specific
network dynamics defined by the combination of gene positions and cell
growth state.

\emph{Role of noise. }
%
Some relevant considerations can be made about the possible role of
noise in the results given here within a purely deterministic
framework. In general, noise can strongly affect the dynamics of a
repressilator.  For example, in the \textit{in vivo} realization of
the repressilator~\cite{Elowitz2000} only about $40\%$ of the cells
displayed oscillations, with high cell-to-cell variability in
oscillation period and amplitude. Several studies have analyzed the
possible impact of noise, focusing on the stochasticity that arises
from the discrete nature of the molecular players and from the
inherent randomness of their interactions (together referred to as
\textit{intrinsic noise})~\cite{Elowitz2000,Strelkowa2010,Yoda2007}.
The main result is that intrinsic noise can both play a constructive
role in oscillation robustness and a destructive one. 
The constructive phenomenon can enlarge the parameter space of
oscillations through a resonance effect~\cite{Yoda2007}. The destructive one 
causes strong cell-to-cell variability in oscillation amplitude and 
period~\cite{Elowitz2000}.  

While intrinsic noise can be a relevant factor and could partially
explain the experimentally observed variability, the dominant source
of noise might be due to fluctuations in global cellular parameters
(\textit{extrinsic noise}), such as ribosome or polymerases
concentration. This has been shown to be the case in \textit{E.~coli},
for relatively high expression (more than approximately $10$ proteins
per cell) ~\cite{Taniguchi2010}.
Since oscillations in the repressilator generally require strong
promoters \rev{(high basal expressions)}, the variability in the circuit
dynamics is expected to be highly sensitive to the extrinsic noise
level, which adds up, for plasmids, to the aforementioned cell-to-cell
copy-number variation.
These considerations point to an interest in considering the
stochastic aspects of the circuit. However, in order to
extend the mean-field model introduced here to analyze the growth rate
dependence of the cell-to-cell variability of the repressilator
dynamics, it would be necessary to know how the extrinsic noise scales
with growth rate.  Unfortunately, there are no experimental data
concerning this scaling, making the extension of this work to the
stochastic case premature.

Nevertheless, the possibility of a switch between different dynamical
regimes in response to the physiological cell state opens interesting
considerations about the robustness of this oscillatory genetic
circuit. Fluctuations in physiological parameters such as the growth
rate fall in the broadly-defined category of extrinsic noise.  It has
been suggested that fluctuations in the growth rate, mainly through
its influence on protein dilution, can account for a considerable part
of the measured extrinsic noise~\cite{Marathe2012,Tsuru2009}.  Our
analysis suggests that fluctuations in cell parameters linked to
growth can introduce cell-to-cell variability in the circuit dynamics
(convergence to oscillation or to a stable fixed point) as well as in
oscillation period and amplitude. Therefore, the reasons behind the
lack of robustness of the repressilator realized \textit{in
  vivo}~\cite{Elowitz2000} should be also searched in the variability
of physiological parameters rather than focusing exclusively on
intrinsic noise effects.

\emph{\rev{Biological} outlook. } Finally, we believe these findings
could be relevant from both a systems biology and a synthetic biology
perspective.
There is a long list of endogenous oscillators in
bacteria~\cite{Lenz2011} that are interesting for the former
discipline, and need to be understood within the framework adopted
here.
The most important examples are circadian clocks and the cell cycle
itself.
We previously studied the dynamics of the DnaA\footnote{DnaA is a protein responsible for the initiation of DNA replication in several bacteria. 
Essentially, if present in a sufficiently high concentration, it can promote the melting of DNA strands at the replication origin.} oscillatory circuit,
which is determinant in this last process~\cite{Grant2011}.  In this
case, the timescale of oscillations matches (by definition) the cell
cycle time, thus the approximations defined in
section~\ref{growth-rate-dep} are not valid. More complex models are
required, and there is no commonly accepted \rev{theoretical
  framework to describe this case}.  However, it is interesting to note
that in the simple modeling framework adopted here, the period of
oscillation decreases with growth rate without the need of specific
additional regulation. In absence of overlapping replication rounds,
this is exactly the kind of behavior desired for an
oscillator regulating the triggering of DNA replication, such as the
DnaA circuit: a shortening of the initiation time is required when the
cell volume grows faster to synchronize DNA replication and volume
doubling (the situation becomes more complex at fast
growth~\cite{Grant2011}).

In contrast, circadian clocks need to be resilient to changes in the
cell doubling time, and thus in the growth rate, in order to keep a
steady $24$-hour period in variable environmental conditions, and thus
can not measure time using the cell cycle.  The consequent decoupling
between the cell cycle and the circadian rhythm has indeed been
verified in cyanobacteria~\cite{Kondo1997,Mori2001,Mihalcescu2004}.
Our results suggest that the dynamics of a genetic oscillator is 
naturally strongly connected to the cellular growth rate.  Therefore,
specific regulatory mechanisms are required in order to compensate for
these effects and render a circadian oscillator insulated from the
growth state.  Although circadian clocks appear to be primarily based
on post-translational circuits in bacteria~\cite{Lenz2011}, the
proteins involved are the result of a gene expression process, and
thus in principle coupled with growth rate~\cite{Klumpp2009}.  
It would be interesting to evaluate experimentally if the promoters
regulating these proteins are more buffered as a function of growth rate
compared to others.
Quantitative models taking into account the parameter dependence on
growth rate, such as the one presented here, may be important to pose
the question of the growth-rate robustness of the circadian cycle.
For example, the circuit architectures and the type of regulations
selected by evolution to compose circadian oscillators might be, at
least in part, constrained by the implementation of the observed
growth-rate independence.

Finally, from a synthetic biology standpoint, changing the conditions
in which the cells are grown alters quantitatively the characteristics
of the repressilator dynamics in a predictable manner. This offers the
possibility of external control of the circuit behavior by simply
operating on macroscopic variable related to physiology such as the
type of nutrient supply or the temperature. This way, the engineering
and control of the dynamics can be performed by tuning environmental
conditions in a model-guided way, rather then by modifying the genetic
components, which can be technically complex. 
%

\section*{Acknowledgements}

We acknowledge useful discussions with Vittore F. Scolari, Mina Zarei
and Bianca Sclavi. We thank Matthew AA Grant and Carla Bosia for
critical reading of the manuscript.  This work was supported by the
the International Human Frontier Science Program Organization (Grant
RGY0069/2009-C).

\appendix

\section{Population averages vs cell-cycle averages}
\label{app1}

A common assumption in genetic circuit modeling is that the
contribution of the cell-cycle dynamics can be neglected, at least
when one is interested in time scales longer than the doubling time,
which is often the case given the typical high protein stability. This
approximation allows the use of effective parameters obtained
averaging over the cell cycle. \rev{The contribution of protein
  dilution due to growth and division is incorporated as an effective
  degradation rate}~\cite{Marathe2012}.
However, as discussed in the text, the growth-rate dependences of gene
expression parameters are derived from experimental observations of
average cellular properties in a population~\cite{Bremer1996}, which
are also affected by the cell age distribution.  \rev{A quantitative
  estimate of the age-structure effects on measurements performed on
  an exponentially growing population is especially important when
  their growth-rate dependence is in analysis. In fact, the fraction
  of cells found at a certain cell-cycle stage is itself a function of
  the growth rate.}

This Appendix \rev{discusses} the quantitative difference between
population and cell-cycle averages for two quantities that are well
characterized, the gene dosage due to DNA replication and the cellular
volume. We will show that in these two cases population averages can
be used in dynamic models of genetic circuits without introducing
significant errors. The assumption that this result can be generalized
to other quantities justifies, although not rigorously, the use of
available experimentally estimated population averages in dynamic
models for those quantities whose time dependence (or even cell-cycle
averages) are not known~\cite{Klumpp2009}. This is the case for
ribosome or polymerase concentrations at different growth rates, which
are crucial to determine the growth-rate dependence of transcription
and translation rates in Eq.~\ref{constitutive}.

\rev{Thus, in our case, given the small quantitative difference
  between the two type of averages, the empirical growth rate
  dependences (based on population averages) have been used in the
  analysis (and thus for all the parameters quoted in the main
  text). The alternative use of cell-cycle averages is possible only
  for the gene dosage and volume, and does not alter significantly the
  results. Indeed, as reported in~\cite{Marathe2012}, for a
  constitutive gene the average protein concentrations calculated with
  respect to an age-structured population or with respect to the cell
  cycle differ by only a few percent.}

\subsection{Cooper-Helmstetter model and gene dosage}
\label{CH-model}

\begin{figure}[t]
\begin{center}
\epsfig{file=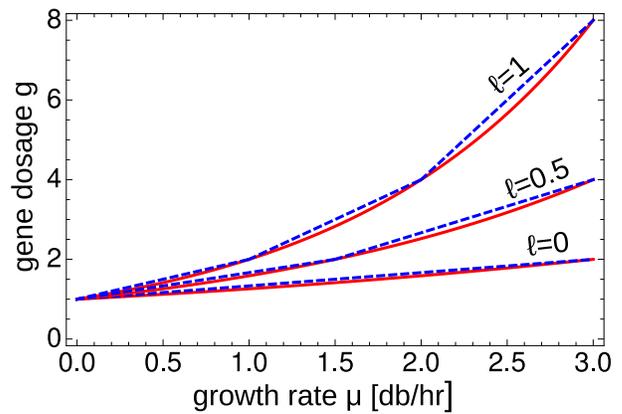, width=8cm}\\
\caption{(Color online) \textbf{Cell cycle and population average of
    gene dosage.} The gene dosage $g$ averaged over the cell cycle
  (dashed blue lines) and averaged over a population in balanced
  exponential growth (continuous red lines) are shown as a function of
  the growth rate for different chromosomal positions $\ell$
  (normalized distance from Ori). In the physiological range of growth
  rates, the two quantities do not present a substantial quantitative
  difference.\label{figapp}}
\end{center}
\end{figure}

DNA replication in fast-growing bacteria such as \textit{E.~coli}
typically starts from a single replication origin (Ori) and proceeds
bidirectionally along the circular chromosome until it reaches the
replication terminus (Ter).  The Cooper-Helmstetter
model~\cite{Cooper1968} establishes the relation between growth rate
and replication timing such that DNA copies are produced on time for
each newborn cell.  The model is based on the empirical observation
that the time necessary for chromosome replication (called ``$C$
period'') and the time period between completion of chromosome
replication and the following cell division ($D$ period) are
approximately constant (at least for fast-dividing
cells~\cite{Michelsen2003}, with doubling times less than 1hr). Since
at time $C+D$ the cell divides, a time lag $X$ before initiation is
necessary to make the total replication time $X+C+D$ an integer
multiple of the doubling time $\tau$, ``synchronizing'' DNA
replication and cell division. Thus, the following relation has to be
satisfied
\begin{equation}
X+C+D =(n+1)\tau \ ,
\end{equation}
where $n=Int[\frac{C+D}{\tau}]$ is the integer number of times that
$\tau$ divides $C+D$.  Starting from this relation, it can be easily
shown that the number of origins present at initiation is exactly
$2^n$~\cite{Cooper1968,Bremer1996}.  More generally, we can consider a
gene at a chromosomal position defined by its normalized distance $\ell$
from Ori, i.e. $\ell=0$ represents a gene in Ori and $\ell=1$ in Ter.  The
copy number of this gene, $g$, changes during the cell cycle
following
\begin{displaymath}
g(t) = \left\{ \begin{array}{ll}
2^{n'} & ~\textrm{if~ $0 <t < (n'+1)\tau -(C(1-\ell)+D)$}\\
2^{n'+1} &~ \textrm{if~ $ (n'+1)\tau -(C(1-\ell)+D)< t< \tau$}
\end{array} \right.  \ ,
\label{origin}
\end{displaymath}
where $n'=Int[\frac{C(1-\ell)+D}{\tau}]$. Therefore, the gene dosage
averaged over the cell cycle (which could be measured following a
single cell lineage and averaging over time), is given by
\begin{eqnarray}
&&\langle g(t)\rangle_{cell~cycle} = \frac{1}{\tau} \int_{0}^{\tau}
g(t) dt=\nonumber\\ 
&&= 2^{n'} [1 -n' + \mu (C(1-\ell)+D)]  \ .
\end{eqnarray}

On the other hand, the population age structure must be taken into
account when evaluating the average gene dosage in a cell population.
For ideal ``balanced exponential'' growth with rate $\mu$ this
distribution is given by~\cite{Powell1956,Marathe2012}
\begin{equation}
 a(t,\mu)= 2~ln2~\mu ~2^{-\mu t} .
\end{equation}
Thus, the population average is
\begin{eqnarray}
\langle g(t)\rangle_{population} &=& \int_{0}^{\tau} a(t,\mu) g(t)dt=\nonumber\\
&=& 2^{\mu [C(1-\ell) + D]} \ .
\end{eqnarray}
This is the expression typically used to evaluate the gene
dosage~\cite{Klumpp2009,Bremer1996}. As shown in Fig.~\ref{figapp},
the difference between cell-cycle averages (dashed blue lines) and
population averages (red continuous lines) in the physiological range
of growth rates is negligible for all gene positions.

\subsection{Cell volume growth}

Similar considerations can be carried out in the case of the average
cell volume. The functional form of the volume increase in time during
a cell cycle has long been debated~\cite{Koch1993}, with two
prevailing hypotheses of linear growth (constant rate) or exponential
growth (size-dependent rate), although more complex dependences have
been proposed~\cite{Reshes2008}.  Recent experiments strongly suggest
an exponential growth~\cite{Godin2010}, and we assumed this functional
form (note that the same reasoning could be applied to linear growth
straightforwardly, so this choice has no consequences on any of the
results).  With a volume growth of the form $V(t)=V_0 2^{\mu t}$, the
mean volume over a cell cycle is $\langle V(t)\rangle_{cell~cycle} =
V_0 /ln2$, while the integration over the population leads to $\langle
V(t)\rangle_{population} = 2~ ln2~V_0$.  All the volume growth-rate
dependence is hidden in $V_0$, and experimental results indicate that
this dependence is approximately exponential~\cite{Schaechter1958}.
\rev{For all the situations considered in our study, we verified that
  the different numerical factors introduced by averaging over the
  cell cycle or over the population do not affect significantly the
  growth-rate dependence of the mean volume.}


\end{document}